\documentclass[showpacs,preprintnumbers,amsmath,amssymb,twocolumn,superscriptaddress,prb]{revtex4}
\usepackage{graphicx}
\usepackage{amsfonts}
\usepackage{amsmath, amsthm, amssymb}
\usepackage{lmodern}
\usepackage[usenames]{color} 
\usepackage[normalem]{ulem}

\def\i{\mathrm{i}}     
\newcommand{\e}[1]{\mathrm{e}^{#1}}

\newcommand{\vvr}{\mathbf{r}}
\newcommand{\vzero}{\mathbf{0}}

\newcommand{\ie}{i.e.}

\begin{document}
\title{Quantum criticality in a dissipative (2+1)-dimensional $XY$ model of circulating currents in high-$T_\text{c}$ cuprates}
\author{Iver Bakken Sperstad}
\affiliation{Department of Physics, Norwegian University of
Science and Technology, N-7491 Trondheim, Norway}
\author{Einar B. Stiansen}
\affiliation{Department of Physics, Norwegian University of
Science and Technology, N-7491 Trondheim, Norway}
\author{Asle Sudb{\o}}
\affiliation{Department of Physics, Norwegian University of
Science and Technology, N-7491 Trondheim, Norway}

\date{Received \today}
\begin{abstract}
We present large-scale Monte Carlo results for the dynamical critical exponent $z$ and the spatio-temporal two-point 
correlation function of a (2+1)-dimensional quantum $XY$ model with bond dissipation, proposed to describe
a quantum critical point in high-$T_\text{c}$ cuprates near optimal doping. The phase variables of the model, originating with 
a parametrization of circulating currents within the CuO$_2$ unit cells in cuprates, are compact, 
$\{ \theta_{\vvr,\tau} \}$ $\in [-\pi,\pi \rangle$. The dynamical critical exponent is found to be $z \approx 1$, 
and the spatio-temporal correlation functions are explicitly demonstrated to be isotropic in space-imaginary 
time. 
The model thus has a fluctuation spectrum where momentum and frequency enter on equal footing, rather than having 
the essentially momentum-independent marginal Fermi liquid-like fluctuation spectrum previously reported for 
the same model.
\noindent
\end{abstract}	

\maketitle

Quantum critical points describe systems with diverging length scales at zero temperature, and 
have come into much focus in recent years as possible descriptions of anomalous phenomena in strongly correlated 
fermion systems and systems with competing orders.\cite{Lohneysen-Vojta_RMP} 
One prime example of this is represented by the high-$T_\text{c}$ 
superconducting cuprates, where various types of quantum critical phenomena have been proposed as essential for 
understanding the many unusual normal-state transport properties these systems exhibit. This has, over the past 
quarter of a century, represented one of the major challenges in condensed-matter physics.\cite{Zaanen_history_HTc}  

One successful phenomenological framework 
is to describe the the normal phase around optimal doping as a marginal Fermi 
liquid (MFL),\cite{Varma_MFL} the weakest possible violation of having a non-zero quasiparticle residue at the 
Fermi surface. Among the merits of the MFL phenomenology is that it describes transport properties in this strange 
metallic phase in good accordance with experiments. This follows naturally from the essentially momentum-independent, 
linear-in-frequency, fluctuation spectrum of the MFL hypothesis.\cite{Varma_MFL}

More recent works have pursued a more microscopic foundation of MFL.
The underlying picture is that there exists a quantum critical point (QCP) 
residing at $T=0$ beneath the superconducting dome.\cite{Varma_NFL} The degrees of freedom associated with this QCP are 
circulating currents within the unit cells of the CuO$_2$ layers. The main idea is 
that the MFL phenomenology arises from the quantum critical fluctuations of these currents above the QCP at $T > T_\text{c}$.
It has also been demonstrated how the same fluctuations may give rise to $d$-wave high-$T_\text{c}$ superconductivity.\cite{Aji-Shekhter-Varma_d-wave_SC}
The ordering of such circulating currents upon lowering the temperature from the strange metal region into the pseudogap 
region is a candidate for a possible competing order in this part of the phase diagram.\cite{Varma_theory_pseudogap}
Magnetic order conforming with such circulating currents has in fact been observed in several experiments.
\cite{Fauque_orbital_currents, Li_magnetic_order, Baledent-Fauque_orbital_order, Li_collective_modes, Kaminski_TRSB} 
It must be mentioned that others argue that such signatures may have a quite different origin,
\cite{Sonier_no_orbital_currents, Strassle_no_orbital_currents,Borisenko_TRSB,MacDougall_TRSB} 
and also numerical results disagree on the presence of such circulating 
currents,\cite{Weber-Giamarchi_orbital_currents,Greiter-Thomale_orbital_currents,Nishimoto_orbital_currents}
but the model remains one of the central theories of the physics of high-$T_\text{c}$ cuprates.\cite{Vojta_stripe_review,Zaanen_history_HTc}

A remarkable implication of a $q$-independent fluctuation spectrum, such as that posited in MFL theory, is that the 
associated QCP exhibits \emph{local} quantum criticality (LQC). Defining the dynamical critical exponent $z$ from the 
scaling of momentum and frequency at the quantum critical point, $\omega \sim q^z$, this means that, formally, 
$z = \infty$. It is a highly nontrivial question as to how such a remarkable property of a quantum critical point can arise 
in an extended system. Recently, it was argued\cite{Aji-Varma_orbital_currents_PRL, Aji-Varma_dissipative_XY} that 
precisely such local criticality is found in a (2+1)-dimensional quantum $XY$ model with bond dissipation of the 
Caldeira-Leggett\cite{Caldeira-Leggett} form. The angle variables of this model were associated with circulating current 
degrees of freedom, as will be explained below.
 
The results of Ref. ~\onlinecite{Aji-Varma_orbital_currents_PRL} would imply that the previously hypothesized MFL 
fluctuation spectrum has been derived from a microscopic theory applicable to cuprates. In a broader perspective, 
it is of considerable interest to investigate in detail if such unusual behavior can occur in model systems of 
condensed matter, as related variants of locality have also been considered in the context of gauge/gravity 
duality\cite{Faulkner_strange_metal} and QCPs in disordered systems 
and heavy fermion compounds.\cite{Lohneysen-Vojta_RMP}

From naive scaling arguments\cite{Hertz_quantum_critical, Sperstad_dissipative} applied to the 
dissipative model proposed in Ref. ~\onlinecite{Aji-Varma_orbital_currents_PRL}, one might expect that 
dissipation is irrelevant in the renormalization group sense. The result would then not be LQC, but instead 
conventional quantum criticality with isotropic scaling, $z = 1$. Here, we report results from Monte Carlo 
simulations performed directly on the (2+1)-dimensional quantum $XY$ model with bond dissipation and compact 
angle variables, considered in Ref. ~\onlinecite{Aji-Varma_orbital_currents_PRL}. Our results strongly indicate 
that in this model, $z=1$.

The dissipative (2+1)-dimensional [(2+1)D] $XY$ action considered in Ref. ~\onlinecite{Aji-Varma_orbital_currents_PRL} takes the form
\begin{align}
  \label{eq:action}
  S &= -K \sum_{\langle \vvr, \vvr' \rangle } \sum_{\tau=1}^{L_\tau}\cos( \Delta\theta_{\vvr,\vvr',\tau} ) \\ \nonumber
  &-K_\tau \sum_{\vvr} \sum_{\tau=1}^{L_\tau}\cos(\theta_{\vvr,\tau + 1}- \theta_{\vvr,\tau}) \\ \nonumber
  &+ \frac{\alpha}{2}\sum_{\langle \vvr, \vvr' \rangle} \sum_{\tau\neq\tau'}^{L_\tau}\left(\frac{\pi}{L_\tau}\right)^2
\frac{\left(\Delta\theta_{ \vvr,\vvr',\tau} - \Delta\theta_{\vvr,\vvr',\tau'}\right)^2}{\sin^2(\frac{\pi}{L_\tau}|\tau-\tau'|)},
\end{align}
when put on a cubic $L \times L \times L_\tau$ lattice. The bond variables are given by 
$\Delta\theta_{\vvr,\vvr',\tau} = \theta_{\vvr,\tau} - \theta_{\vvr',\tau}$, where the sum over $\vvr$ and $\vvr'$ goes 
over nearest neighbors in the $x$-$y$ plane. Periodic boundary conditions are implicit in the imaginary time 
direction, and are also applied in the spatial directions.

Such a model has previously been employed as an effective description of a resistively shunted Josephson junction 
array,\cite{Chakravarty_dissipative_PT} and it may also be viewed as a generic quantum rotor 
model with dissipative currents.
A third possible interpretation in the context of high-$T_\text{c}$ cuprates is as follows. Suppose the angles $\theta$ 
{\it a priori} can take only four possible values. These four values then represent the directions of a 
pseudospin associated with the four possible ordered circulating current patterns within each CuO$_2$ unit cell  (see, e.g., Fig. 1 
of Ref. ~\onlinecite{Borkje_orbital_currents}). The first two terms represent the standard interaction energies in 
space-imaginary time of these circulating currents in neighboring unit cells, and have been 
derived from microscopics.\cite{Borkje_orbital_currents} The last term is the term responsible for dissipating 
the ordered circulating currents.\cite{Aji-Varma_orbital_currents_PRL}

In Eq. \eqref{eq:action}, the angles are continuous variables. 
We will discuss a possible {\it a posteriori} 
justification for this later in this Rapid Communication, by showing that an added four-fold anisotropy term is perturbatively irrelevant. 
Reference ~\onlinecite{Aji-Varma_orbital_currents_PRL}, moreover, appears to treat $\theta_{\vvr,\tau}$ as 
compact variables, also in the presence of a dissipation term that apparently renders the action nonperiodic 
in the angle variables.\cite{Schon-Zaikin} 
In order to investigate numerically the same model 
considered in Ref. ~\onlinecite{Aji-Varma_orbital_currents_PRL}, we therefore \emph{compactify} the expression 
$\Delta \theta_{\vvr,\vvr',\tau} - \Delta \theta_{\vvr,\vvr',\tau'}$ so that it is defined modulo $2\pi$.
We will discuss alternative choices later.

The calculations of Ref. ~\onlinecite{Aji-Varma_orbital_currents_PRL} were not restricted to any specific parameter regime, 
but predicted that every point on the $T=0$ quantum critical surface in $\alpha-K-K_\tau$ (parameter) space
(for $\alpha > 0$) should 
be a local QCP. Accordingly, we choose convenient coupling constants when searching for LQC in our 
simulations, and for the results presented here, the dissipation strength is fixed at $\alpha = 0.05$.

The phase diagram (not shown) is qualitatively very similar to those
found for related compact (1+1)D models with bond dissipation.\cite{Sperstad_dissipative,Stiansen_compactness}
It features a single critical surface that separates a disordered from a fully ordered phase, 
and which is continuously connected to the 3D $XY$ critical line at $\alpha = 0$. For similar models in (1+1) 
dimensions, only the region of relatively moderate dissipation was accessible to simulations, as increasing 
$\alpha$ increases finite-size effects, resulting in apparent values $z < 1$ for the dynamical critical exponent. 
As expected, this problem is no less severe in (2+1) dimensions. Available system sizes are restricted by the 
absence of cluster algorithms to treat models with bond dissipation appropriately,\cite{Sperstad_dissipative} 
and we are therefore confined to local Metropolis updates. 

To locate the phase transition, we vary the spatial coupling $K$ and use the crossing point for different system sizes 
$L$ of the Binder cumulant 
$g = 1 - \langle |m|^4  \rangle / (2 \langle |m|^2 \rangle^2)$.
Here, $m = \sum_{\vvr,\tau}{\exp{\lbrack \i \theta_{\vvr,\tau} \rbrack}}$ is the order parameter of 
the $U(1)$-symmetric degrees of freedom. Due to the anisotropy of the interactions, we have to calculate $g$ for 
multiple values of $L_\tau$ for each spatial system size $L$, as described in more detail, e.g., in Ref. 
~\onlinecite{Sperstad_dissipative}. The value $L_\tau = L_\tau^\ast$ where the function $g(L_\tau)$ reaches its 
maximum corresponds to the \emph{optimal} temporal extent for which the system appears as isotropic as it can be, 
the anisotropic interactions taken into account. 

For a conventional QCP, at which the correlation length $\xi_\tau$ in imaginary time scales with the correlation length 
$\xi$ in space as $\xi_\tau \sim \xi^z$ with a finite $z$, we expect to observe the scaling relation $L_\tau^\ast \sim L^z$. 
This scaling procedure then allows one to extract the dynamical critical exponent $z$ from Binder cumulant data. 
For a local QCP formally having $z = \infty$, we expect this scaling to break down. Our strategy to search for possible 
LQC in the model \eqref{eq:action} is therefore to perform the above procedure \emph{assuming} conventional criticality, 
and then look for indications that this hypothesis should be rejected.

\begin{figure}[h!]
	\includegraphics[width=0.45\textwidth]{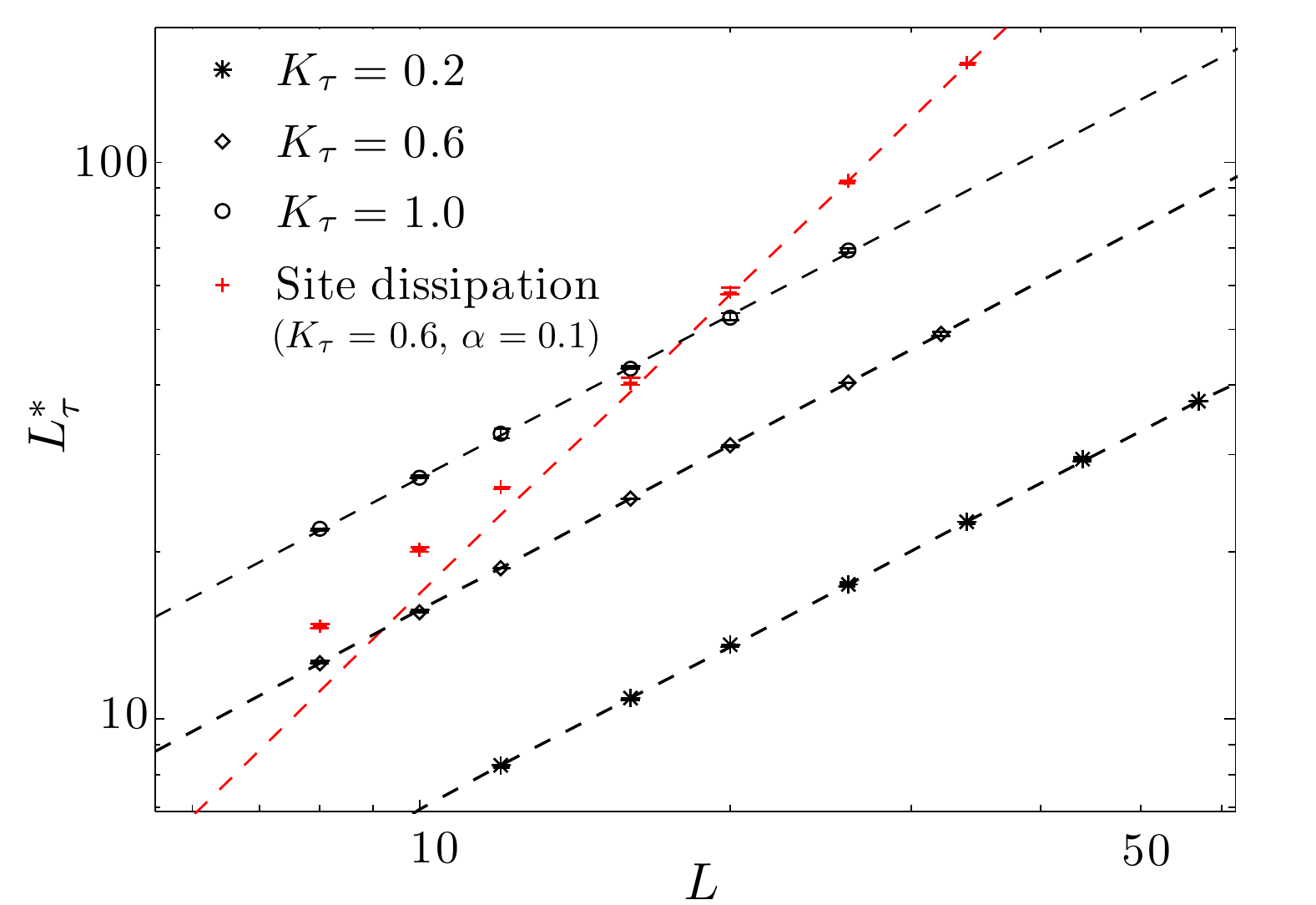}
	\caption{(Color online) Finite-size analysis of the maximum $L_\tau^\ast$ of the Binder cumulant curves $g(L_\tau)$ as a function 
	of spatial system size $L$. For the black data points, the dynamical critical exponents $z$ as given in 
	Table \ref{tab:exponents} are obtained from the slope of the fitting lines (dashed). 
	The red (gray) points show similar results for site dissipation for comparison, 
	where a fit of the three largest systems yields $z=1.84(3)$.} 
	\label{fig:Lz_vs_L_bond-site}
\end{figure}

The results of this finite-size analysis is shown in Fig. \ref{fig:Lz_vs_L_bond-site}, with the values of the 
dynamical critical exponent $z$ given in Table \ref{tab:exponents}. 
Here, we have chosen three different values of the 
quantum coupling $K_\tau$ in order to investigate both the limit of relatively weak quantum coupling 
and the opposite limit leading to relatively strong system anisotropy.

\begin{table}
	\caption{Critical coupling $K_\text{c}$ and dynamical critical exponent $z$ for different values of the quantum coupling 
	$K_\tau$, but for the same dissipation strength $\alpha = 0.05$. 
	Uncertainty estimates for $z$ have 
	been calculated by a bootstrap procedure, including the uncertainty in $K_\text{c}$.}
	\begin{center}
    \begin{tabular*}{0.25\textwidth}{@{\extracolsep{\fill}}  c  c  c}
    \hline
    \hline    
    $K_\tau$ & $K_\text{c}$ & $z$ \\ 
    \hline
    0.2 	& 0.48068(5) & 0.968(8) \\
    0.6 	& 0.28244(4) & 0.985(8) \\ 
    1.0 	& 0.18008(5) & 0.970(11) \\ 
    \hline
    \hline
    \end{tabular*}
    \label{tab:exponents}
	\end{center}
\end{table}

The results show that the effective dynamical critical exponent is $z \lesssim 1$ for all the parameter sets 
considered, and we expect that we could obtain $z \approx 1$ if we were able to reach higher values of $L$. 
(For a smaller value $\alpha = 0.02$, we obtained $z=1$ within statistical uncertainty.)
It is conceivable that signatures of LQC would be visible only for systems 
larger than the admittedly moderate system sizes accessible to present algorithms. 
However, were that the case, the true $z \to \infty$ nature of the model would likely reveal itself 
as strongly increasing effective values of $z$ as a finite-size effect for increasing $L$. For comparison, we 
have also carried out simulations with equivalent parameters of a (2+1)D  $XY$ model with \emph{site} 
dissipation, for which $z=2$ is expected. 
\cite{Werner-Troyer-Sachdev_dissipative_XY_chain, Sperstad_dissipative}
The results are included in Fig. \ref{fig:Lz_vs_L_bond-site}, 
and already for system sizes comparable to those for bond dissipation, we observe (finite-size) crossover 
behavior with $z \to 2$.
For bond dissipation, we observe no tendency toward $z>1$ for 
either of the parameter sets, and it is hard to imagine how crossover to $z \to \infty$ scaling should 
be much slower than crossover to $z = 2$ scaling.

For all results reported here, we have used parallel tempering\cite{Hukushima_parallel_tempering} to reduce 
autocorrelation times, and to ensure that the simulations are well equilibrated. To emulate the continuous 
$U(1)$ symmetry, the simulations are made for $Z_q$ clock models, with $q = 128$ for $K_\tau = 0.2, 1.0$, 
and $q=32$ for $K_\tau = 0.6$. The nature of the criticality remains unchanged also when increasing to 
$q = 1024$. The results are obtained using an implementation of the Mersenne Twister\cite{Mersenne_Twister} 
random number generator, but other random number generators produced consistent results.

\begin{figure}[h!]
	\includegraphics[width=0.45\textwidth]{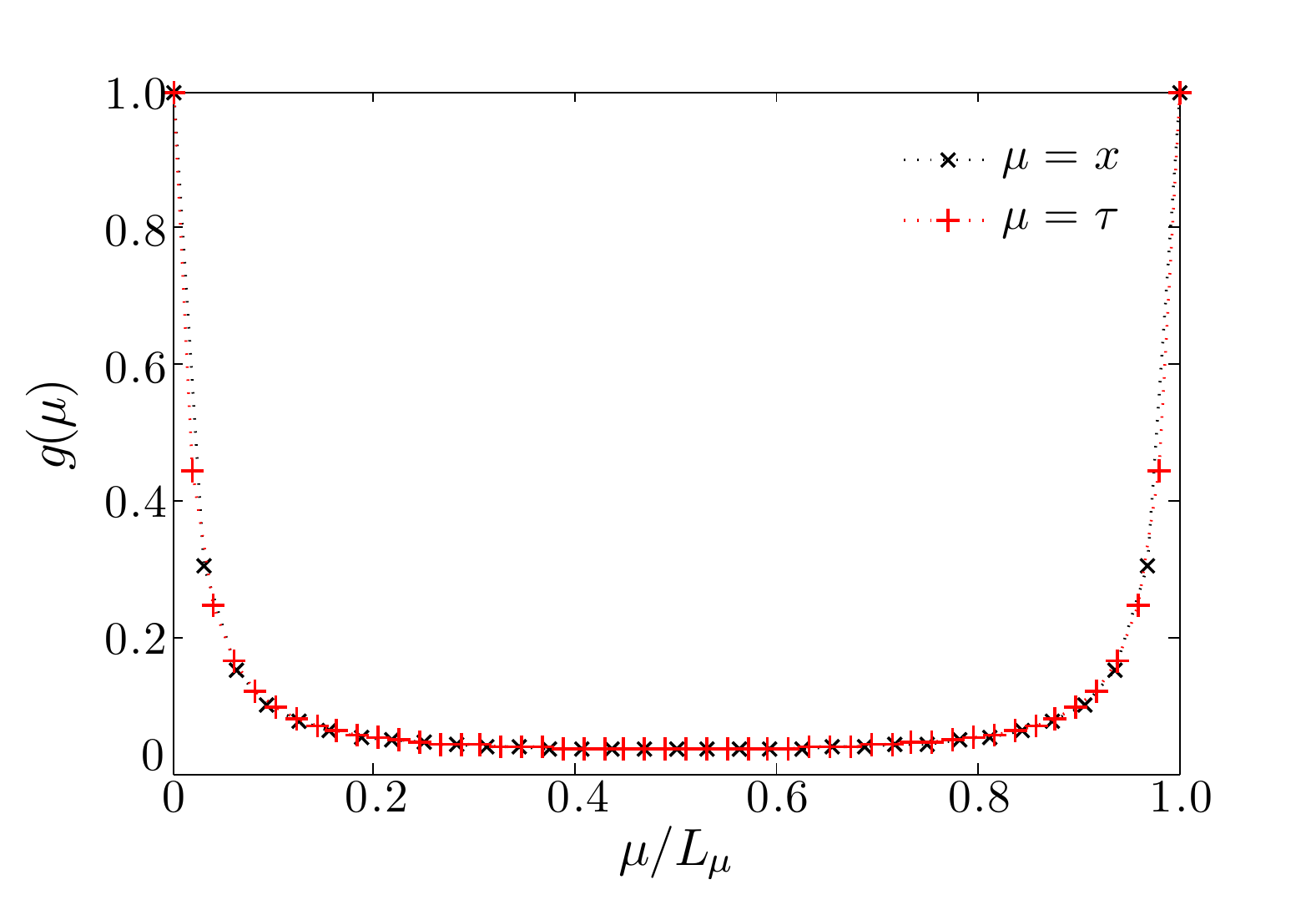}
	\caption{(Color online) Correlation functions at the critical point $K_\text{c} = 0.28244$ for dissipation strength $\alpha = 0.05$ 
	and quantum coupling $K_\tau = 0.6$. The system size $L = 32$, $L_\tau = 49 \approx L_\tau^\ast$ corresponds to the 
	rightmost data point of the midmost data series in Fig. \ref{fig:Lz_vs_L_bond-site}. 
	The correlation function is defined in the spatial direction as 
	$g(x) = g(|\vvr - \vvr'|) = C(\vvr - \vvr',0)$ 
	and in the temporal direction as  $g(\tau) = C(\vzero,\tau)$, with $C$ defined in 
	Eq. \eqref{eq:corrfunc}. Also, $L_x \equiv L$. Error bars are smaller than the line width, 
	and the dotted lines are guides to the eye.} 
	\label{fig:corrfunc}
\end{figure}

Although we found no indication of LQC from the scaling of the Binder cumulant, we also considered the 
correlations of the order parameter field directly,
\begin{equation}
	\label{eq:corrfunc}
	C(\vvr-\vvr',\tau-\tau') = \langle \e{ \i \theta_{\vvr ,\tau} } \e{ -\i \theta_{\vvr',\tau'} }  \rangle.
\end{equation}	
The correlation functions presented here are obtained for the parameter set $K_\tau = 0.6$,
with $L_\tau = L_\tau^\ast$ and $K = K_\text{c}$ as obtained from the previous simulations, and therefore serve as a 
self-consistency check of the Binder scaling procedure. From Fig. \ref{fig:corrfunc}, it is evident that the 
correlation function at the critical point decays isotropically
in space-imaginary time. In other words, there are no signs of locality. 

Furthermore, we have verified that the same conclusion may be drawn for the other values of $K_\tau$ considered, 
and also for larger system sizes with aspect ratios found from extrapolation based on the power law shown in 
Fig. \ref{fig:Lz_vs_L_bond-site}. As an additional test, we compared the correlation functions shown here with 
those obtained by setting $\alpha = 0$ in Eq. \eqref{eq:action}. Letting $K_\tau > K$, values of $L_\tau$ and $K_\text{c}$ 
were determined by the same procedure as for the dissipative model. There is no indication that adding dissipation 
changes the scaling of the temporal correlation length $\xi_\tau$ with respect to the spatial correlation length $\xi$.

Depending on how $\Delta \theta_{\vvr,\vvr',\tau}$ is interpreted in the dissipation term, 
it may be argued either that the correct 
treatment is to compactify only the gradients $\Delta \theta_{\vvr,\vvr',\tau}$, restricting them to the 
interval $\lbrack -\pi, \pi \rangle$, or to do so to the difference 
$\Delta \theta_{\vvr,\vvr',\tau} - \Delta \theta_{\vvr,\vvr',\tau'}$ as well. 
Although we have chosen the latter, as in 
Ref. ~\onlinecite{Stiansen_compactness}, we also performed simulations with the former compactification 
scheme. The results are qualitatively  similar, with the difference merely 
amounting to a renormalization of the dissipative coupling $\alpha$. In other words, the absence of LQC in 
this model is not contingent on the choice of compactification scheme. 

As explained in connection with Eq. \eqref{eq:action}, the underlying circulating 
current degrees of freedom are most naturally described by discrete, $Z_4$-symmetric variables.
In Ref. ~\onlinecite{Aji-Varma_dissipative_XY}, it was argued that a model with continuous $U(1)$ symmetry 
nonetheless would be a correct description. The result of LQC would then also apply to the four-state model 
of the original degrees of freedom since a fourfold anisotropy field, given by 
\begin{equation}
  \label{eq:anisotropy}
  S_4 = h_4 \sum_{\vvr, \tau} \cos{(4 \theta_{\vvr, \tau})},
\end{equation}
would be irrelevant at the critical point of the action \eqref{eq:action}. 
We have investigated the effect of a fourfold anisotropy in our simulations
by including the term \eqref{eq:anisotropy} in the action.
Using the approach of Ref. ~\onlinecite{Caselle-Hasenbusch_cubic_fixed_point}, we find the same result 
for the dissipative (2+1)D $XY$ model as reported there for the classical 3D $XY$ model, namely, that 
the $h_4$ term is perturbatively irrelevant.

The soft constraint represented by a (finite) anisotropy term is not obviously the same as the hard 
constraint constituted by the discrete $Z_4$ variables of the original model (the limit $h_4 = \infty$). 
We may only speculate whether a putative LQC fixed point for a $U(1)$ theory might survive in the 
limit $h_4 \to \infty$, but note that our simulations showed no signs of locality neither when enforcing 
a soft nor a strong $Z_4$ constraint on the variables.\cite{fotnote}

Finally, we briefly consider variants of LQC other than that of Ref. ~\onlinecite{Aji-Varma_dissipative_XY},
which predicts a strictly infinite $z$ for $\xi_\tau \sim \xi^z$ so that $\xi$ is strictly vanishing at 
criticality. Another conceivable sense in which $z \to \infty$ is by 
activated dynamical scaling, \cite{Vojta_computing_QPT} \ie, scaling on the form $\ln{\xi_\tau} \sim \xi^\psi$.  
In this case, as we expect also in the first case, locality would 
manifest itself in our simulations as a strongly increasing value of $z>1$ as the thermodynamical limit was 
approached. This is not observed in our results. 
We have also verified explicitly, by an appropriate modification of the 
scaling,\cite{Vojta_computing_QPT} that our results are not consistent with activated dynamical scaling. 

In conclusion, we find no signs of local quantum criticality in the compact (2+1)D $XY$ model with bond dissipation, 
but instead conventional quantum criticality with indications of isotropic scaling of imaginary time and space. 
This implies that the fluctuation spectrum of the model is a function of the combination $\sqrt{q^2 + \omega^2}$,
rather than being dependent only on the frequency $\omega$, but not on the momentum $q$ (which would be a
hallmark of local quantum criticality). Our results therefore differ in a fundamental way from those obtained from 
the same model in Ref. \onlinecite{Aji-Varma_orbital_currents_PRL}.

The authors acknowledge useful discussions with Egil V. Herland, Chandra M. Varma, and Vivek Aji. 
A.S. was supported by the Norwegian Research Council under Grant No. 167498/V30  (STORFORSK). E.B.S. and 
I.B.S thank NTNU for financial support. The work was also supported through the Norwegian consortium
for high-performance computing (NOTUR).


\begin{thebibliography}{35}

\expandafter\ifx\csname natexlab\endcsname\relax\def\natexlab#1{#1}\fi
\expandafter\ifx\csname bibnamefont\endcsname\relax
  \def\bibnamefont#1{#1}\fi
\expandafter\ifx\csname bibfnamefont\endcsname\relax
  \def\bibfnamefont#1{#1}\fi
\expandafter\ifx\csname citenamefont\endcsname\relax
  \def\citenamefont#1{#1}\fi
\expandafter\ifx\csname url\endcsname\relax
  \def\url#1{\texttt{#1}}\fi
\expandafter\ifx\csname urlprefix\endcsname\relax\def\urlprefix{URL }\fi
\providecommand{\bibinfo}[2]{#2}
\providecommand{\eprint}[2][]{\url{#2}}

\bibitem[{\citenamefont{v.~L\"{o}hneysen
  et~al.}(2007)\citenamefont{v.~L\"{o}hneysen, Rosch, Vojta, and
  W\"{o}lfle}}]{Lohneysen-Vojta_RMP}
\bibinfo{author}{\bibfnamefont{H.}~\bibnamefont{v.~L\"{o}hneysen}},
  \bibinfo{author}{\bibfnamefont{A.}~\bibnamefont{Rosch}},
  \bibinfo{author}{\bibfnamefont{M.}~\bibnamefont{Vojta}}, \bibnamefont{and}
  \bibinfo{author}{\bibfnamefont{P.}~\bibnamefont{W\"{o}lfle}},
  \bibinfo{journal}{Rev. Mod. Phys.} \textbf{\bibinfo{volume}{79}},
  \bibinfo{pages}{1015} (\bibinfo{year}{2007}).

\bibitem[{\citenamefont{Zaanen}(2011)}]{Zaanen_history_HTc}
\bibinfo{author}{\bibfnamefont{J.}~\bibnamefont{Zaanen}}, in
  \emph{\bibinfo{booktitle}{100 years of superconductivity}}, edited by
  \bibinfo{editor}{\bibfnamefont{H.}~\bibnamefont{Rogalla}} \bibnamefont{and}
  \bibinfo{editor}{\bibfnamefont{P.~H.} \bibnamefont{Kes}}
  (\bibinfo{publisher}{Taylor \& Francis, London}, \bibinfo{year}{2011})
  \bibinfo{note}{(in press) (e-print arXiv:1012.5461)}.

\bibitem[{\citenamefont{Varma et~al.}(1989)\citenamefont{Varma, Littlewood,
  Schmitt-Rink, Abrahams, and Ruckenstein}}]{Varma_MFL}
\bibinfo{author}{\bibfnamefont{C.~M.} \bibnamefont{Varma}},
  \bibinfo{author}{\bibfnamefont{P.~B.} \bibnamefont{Littlewood}},
  \bibinfo{author}{\bibfnamefont{S.}~\bibnamefont{Schmitt-Rink}},
  \bibinfo{author}{\bibfnamefont{E.}~\bibnamefont{Abrahams}}, \bibnamefont{and}
  \bibinfo{author}{\bibfnamefont{A.~E.} \bibnamefont{Ruckenstein}},
  \bibinfo{journal}{Phys. Rev. Lett.} \textbf{\bibinfo{volume}{63}},
  \bibinfo{pages}{1996} (\bibinfo{year}{1989}).

\bibitem[{\citenamefont{Varma}(1997)}]{Varma_NFL}
\bibinfo{author}{\bibfnamefont{C.~M.} \bibnamefont{Varma}},
  \bibinfo{journal}{Phys. Rev. B} \textbf{\bibinfo{volume}{55}},
  \bibinfo{pages}{14554} (\bibinfo{year}{1997}).

\bibitem[{\citenamefont{Aji et~al.}(2010)\citenamefont{Aji, Shekhter, and
  Varma}}]{Aji-Shekhter-Varma_d-wave_SC}
\bibinfo{author}{\bibfnamefont{V.}~\bibnamefont{Aji}},
  \bibinfo{author}{\bibfnamefont{A.}~\bibnamefont{Shekhter}}, \bibnamefont{and}
  \bibinfo{author}{\bibfnamefont{C.~M.} \bibnamefont{Varma}},
  \bibinfo{journal}{Phys. Rev. B} \textbf{\bibinfo{volume}{81}},
  \bibinfo{pages}{064515} (\bibinfo{year}{2010}).

\bibitem[{\citenamefont{Varma}(2006)}]{Varma_theory_pseudogap}
\bibinfo{author}{\bibfnamefont{C.~M.} \bibnamefont{Varma}},
  \bibinfo{journal}{Phys. Rev. B} \textbf{\bibinfo{volume}{73}},
  \bibinfo{pages}{155113} (\bibinfo{year}{2006}).

\bibitem[{\citenamefont{Fauqu\'e et~al.}(2006)\citenamefont{Fauqu\'e, Sidis,
  Hinkov, Pailh\`es, Lin, Chaud, and Bourges}}]{Fauque_orbital_currents}
\bibinfo{author}{\bibfnamefont{B.}~\bibnamefont{Fauqu\'e}},
  \bibinfo{author}{\bibfnamefont{Y.}~\bibnamefont{Sidis}},
  \bibinfo{author}{\bibfnamefont{V.}~\bibnamefont{Hinkov}},
  \bibinfo{author}{\bibfnamefont{S.}~\bibnamefont{Pailh\`es}},
  \bibinfo{author}{\bibfnamefont{C.~T.} \bibnamefont{Lin}},
  \bibinfo{author}{\bibfnamefont{X.}~\bibnamefont{Chaud}}, \bibnamefont{and}
  \bibinfo{author}{\bibfnamefont{P.}~\bibnamefont{Bourges}},
  \bibinfo{journal}{Phys. Rev. Lett.} \textbf{\bibinfo{volume}{96}},
  \bibinfo{pages}{197001} (\bibinfo{year}{2006}).

\bibitem[{\citenamefont{Kaminski et~al.}(2002)\citenamefont{Kaminski,
  Rosenkranz, Fretwell, Campuzano, Z.~Li, Cullen, You, Olson, Varma, and
  H\"ochst}}]{Kaminski_TRSB}
\bibinfo{author}{\bibfnamefont{A.}~\bibnamefont{Kaminski}},
  \bibinfo{author}{\bibfnamefont{S.}~\bibnamefont{Rosenkranz}},
  \bibinfo{author}{\bibfnamefont{H.~M.} \bibnamefont{Fretwell}},
  \bibinfo{author}{\bibfnamefont{J.~C.} \bibnamefont{Campuzano}},
  \bibinfo{author}{\bibfnamefont{H.~R.} \bibnamefont{Z.~Li}},
  \bibinfo{author}{\bibfnamefont{W.~G.} \bibnamefont{Cullen}},
  \bibinfo{author}{\bibfnamefont{H.}~\bibnamefont{You}},
  \bibinfo{author}{\bibfnamefont{C.~G.} \bibnamefont{Olson}},
  \bibinfo{author}{\bibfnamefont{C.~M.} \bibnamefont{Varma}}, \bibnamefont{and}
  \bibinfo{author}{\bibfnamefont{H.}~\bibnamefont{H\"ochst}},
  \bibinfo{journal}{Nature (London)} \textbf{\bibinfo{volume}{416}},
  \bibinfo{pages}{610} (\bibinfo{year}{2002}).

\bibitem[{\citenamefont{Li et~al.}(2010)\citenamefont{Li, Bal\'edent, Yu,
  Bari\v{s}i\'c, Hradil, Mole, Sidis, Steffens, Zhao, Bourges
  et~al.}}]{Li_collective_modes}
\bibinfo{author}{\bibfnamefont{Y.}~\bibnamefont{Li}},
  \bibinfo{author}{\bibfnamefont{V.}~\bibnamefont{Bal\'edent}},
  \bibinfo{author}{\bibfnamefont{G.}~\bibnamefont{Yu}},
  \bibinfo{author}{\bibfnamefont{N.}~\bibnamefont{Bari\v{s}i\'c}},
  \bibinfo{author}{\bibfnamefont{K.}~\bibnamefont{Hradil}},
  \bibinfo{author}{\bibfnamefont{R.~A.} \bibnamefont{Mole}},
  \bibinfo{author}{\bibfnamefont{Y.}~\bibnamefont{Sidis}},
  \bibinfo{author}{\bibfnamefont{P.}~\bibnamefont{Steffens}},
  \bibinfo{author}{\bibfnamefont{X.}~\bibnamefont{Zhao}},
  \bibinfo{author}{\bibfnamefont{P.}~\bibnamefont{Bourges}},
  \bibnamefont{et~al.}, \bibinfo{journal}{Nature (London)}
  \textbf{\bibinfo{volume}{468}}, \bibinfo{pages}{283} (\bibinfo{year}{2010}).

\bibitem[{\citenamefont{Li et~al.}(2008)\citenamefont{Li, Bal\'edent,
  Bari\v{s}i\'c, Cho, Fauqu\'e, Sidis, Yu, Zhao, Bourges, and
  Greven}}]{Li_magnetic_order}
\bibinfo{author}{\bibfnamefont{Y.}~\bibnamefont{Li}},
  \bibinfo{author}{\bibfnamefont{V.}~\bibnamefont{Bal\'edent}},
  \bibinfo{author}{\bibfnamefont{N.}~\bibnamefont{Bari\v{s}i\'c}},
  \bibinfo{author}{\bibfnamefont{Y.}~\bibnamefont{Cho}},
  \bibinfo{author}{\bibfnamefont{B.}~\bibnamefont{Fauqu\'e}},
  \bibinfo{author}{\bibfnamefont{Y.}~\bibnamefont{Sidis}},
  \bibinfo{author}{\bibfnamefont{G.}~\bibnamefont{Yu}},
  \bibinfo{author}{\bibfnamefont{X.}~\bibnamefont{Zhao}},
  \bibinfo{author}{\bibfnamefont{P.}~\bibnamefont{Bourges}}, \bibnamefont{and}
  \bibinfo{author}{\bibfnamefont{M.}~\bibnamefont{Greven}},
  \bibinfo{journal}{Nature (London)} \textbf{\bibinfo{volume}{455}},
  \bibinfo{pages}{372} (\bibinfo{year}{2008}).

\bibitem[{\citenamefont{Bal\'edent et~al.}(2010)\citenamefont{Bal\'edent,
  Fauqu\'e, Sidis, Christensen, Pailh\`es, Conder, Pomjakushina, Mesot, and
  Bourges}}]{Baledent-Fauque_orbital_order}
\bibinfo{author}{\bibfnamefont{V.}~\bibnamefont{Bal\'edent}},
  \bibinfo{author}{\bibfnamefont{B.}~\bibnamefont{Fauqu\'e}},
  \bibinfo{author}{\bibfnamefont{Y.}~\bibnamefont{Sidis}},
  \bibinfo{author}{\bibfnamefont{N.~B.} \bibnamefont{Christensen}},
  \bibinfo{author}{\bibfnamefont{S.}~\bibnamefont{Pailh\`es}},
  \bibinfo{author}{\bibfnamefont{K.}~\bibnamefont{Conder}},
  \bibinfo{author}{\bibfnamefont{E.}~\bibnamefont{Pomjakushina}},
  \bibinfo{author}{\bibfnamefont{J.}~\bibnamefont{Mesot}}, \bibnamefont{and}
  \bibinfo{author}{\bibfnamefont{P.}~\bibnamefont{Bourges}},
  \bibinfo{journal}{Phys. Rev. Lett.} \textbf{\bibinfo{volume}{105}},
  \bibinfo{pages}{027004} (\bibinfo{year}{2010}).

\bibitem[{\citenamefont{Sonier et~al.}(2009)\citenamefont{Sonier, Pacradouni,
  Sabok-Sayr, Hardy, Bonn, Liang, and Mook}}]{Sonier_no_orbital_currents}
\bibinfo{author}{\bibfnamefont{J.~E.} \bibnamefont{Sonier}},
  \bibinfo{author}{\bibfnamefont{V.}~\bibnamefont{Pacradouni}},
  \bibinfo{author}{\bibfnamefont{S.~A.} \bibnamefont{Sabok-Sayr}},
  \bibinfo{author}{\bibfnamefont{W.~N.} \bibnamefont{Hardy}},
  \bibinfo{author}{\bibfnamefont{D.~A.} \bibnamefont{Bonn}},
  \bibinfo{author}{\bibfnamefont{R.}~\bibnamefont{Liang}}, \bibnamefont{and}
  \bibinfo{author}{\bibfnamefont{H.~A.} \bibnamefont{Mook}},
  \bibinfo{journal}{Phys. Rev. Lett.} \textbf{\bibinfo{volume}{103}}
  (\bibinfo{year}{2009}).

\bibitem[{\citenamefont{Str\"assle et~al.}(2011)\citenamefont{Str\"assle,
  Graneli, Mali, Roos, and Keller}}]{Strassle_no_orbital_currents}
\bibinfo{author}{\bibfnamefont{S.}~\bibnamefont{Str\"assle}},
  \bibinfo{author}{\bibfnamefont{B.}~\bibnamefont{Graneli}},
  \bibinfo{author}{\bibfnamefont{M.}~\bibnamefont{Mali}},
  \bibinfo{author}{\bibfnamefont{J.}~\bibnamefont{Roos}}, \bibnamefont{and}
  \bibinfo{author}{\bibfnamefont{H.}~\bibnamefont{Keller}},
  \bibinfo{journal}{Phys. Rev. Lett.} \textbf{\bibinfo{volume}{106}},
  \bibinfo{pages}{097003} (\bibinfo{year}{2011}).

\bibitem[{\citenamefont{Borisenko et~al.}(2004)\citenamefont{Borisenko,
  Kordyuk, Koitzsch, Kim, Nenkov, Knupfer, Fink, Grazioli, Turchini, and
  Berger}}]{Borisenko_TRSB}
\bibinfo{author}{\bibfnamefont{S.~V.} \bibnamefont{Borisenko}},
  \bibinfo{author}{\bibfnamefont{A.~A.} \bibnamefont{Kordyuk}},
  \bibinfo{author}{\bibfnamefont{A.}~\bibnamefont{Koitzsch}},
  \bibinfo{author}{\bibfnamefont{T.~K.} \bibnamefont{Kim}},
  \bibinfo{author}{\bibfnamefont{K.~A.} \bibnamefont{Nenkov}},
  \bibinfo{author}{\bibfnamefont{M.}~\bibnamefont{Knupfer}},
  \bibinfo{author}{\bibfnamefont{J.}~\bibnamefont{Fink}},
  \bibinfo{author}{\bibfnamefont{C.}~\bibnamefont{Grazioli}},
  \bibinfo{author}{\bibfnamefont{S.}~\bibnamefont{Turchini}}, \bibnamefont{and}
  \bibinfo{author}{\bibfnamefont{H.}~\bibnamefont{Berger}},
  \bibinfo{journal}{Phys. Rev. Lett.} \textbf{\bibinfo{volume}{92}},
  \bibinfo{pages}{207001} (\bibinfo{year}{2004}).

\bibitem[{\citenamefont{MacDougall et~al.}(2008)\citenamefont{MacDougall,
  Aczel, Carlo, Ito, Rodriguez, Russo, Uemura, Wakimoto, and
  Luke}}]{MacDougall_TRSB}
\bibinfo{author}{\bibfnamefont{G.~J.} \bibnamefont{MacDougall}},
  \bibinfo{author}{\bibfnamefont{A.~A.} \bibnamefont{Aczel}},
  \bibinfo{author}{\bibfnamefont{J.~P.} \bibnamefont{Carlo}},
  \bibinfo{author}{\bibfnamefont{T.}~\bibnamefont{Ito}},
  \bibinfo{author}{\bibfnamefont{J.}~\bibnamefont{Rodriguez}},
  \bibinfo{author}{\bibfnamefont{P.~L.} \bibnamefont{Russo}},
  \bibinfo{author}{\bibfnamefont{Y.~J.} \bibnamefont{Uemura}},
  \bibinfo{author}{\bibfnamefont{S.}~\bibnamefont{Wakimoto}}, \bibnamefont{and}
  \bibinfo{author}{\bibfnamefont{G.~M.} \bibnamefont{Luke}},
  \bibinfo{journal}{Phys. Rev. Lett.} \textbf{\bibinfo{volume}{101}},
  \bibinfo{pages}{017001} (\bibinfo{year}{2008}).

\bibitem[{\citenamefont{Greiter and
  Thomale}(2007)}]{Greiter-Thomale_orbital_currents}
\bibinfo{author}{\bibfnamefont{M.}~\bibnamefont{Greiter}} \bibnamefont{and}
  \bibinfo{author}{\bibfnamefont{R.}~\bibnamefont{Thomale}},
  \bibinfo{journal}{Phys. Rev. Lett.} \textbf{\bibinfo{volume}{99}},
  \bibinfo{pages}{027005} (\bibinfo{year}{2007}).

\bibitem[{\citenamefont{Nishimoto et~al.}(2009)\citenamefont{Nishimoto,
  Jeckelmann, and Scalapino}}]{Nishimoto_orbital_currents}
\bibinfo{author}{\bibfnamefont{S.}~\bibnamefont{Nishimoto}},
  \bibinfo{author}{\bibfnamefont{E.}~\bibnamefont{Jeckelmann}},
  \bibnamefont{and} \bibinfo{author}{\bibfnamefont{D.~J.}
  \bibnamefont{Scalapino}}, \bibinfo{journal}{Phys. Rev. B}
  \textbf{\bibinfo{volume}{79}}, \bibinfo{pages}{205115}
  (\bibinfo{year}{2009}).

\bibitem[{\citenamefont{Weber et~al.}(2009)\citenamefont{Weber, L\"auchli,
  Mila, and Giamarchi}}]{Weber-Giamarchi_orbital_currents}
\bibinfo{author}{\bibfnamefont{C.}~\bibnamefont{Weber}},
  \bibinfo{author}{\bibfnamefont{A.}~\bibnamefont{L\"auchli}},
  \bibinfo{author}{\bibfnamefont{F.}~\bibnamefont{Mila}}, \bibnamefont{and}
  \bibinfo{author}{\bibfnamefont{T.}~\bibnamefont{Giamarchi}},
  \bibinfo{journal}{Phys. Rev. Lett.} \textbf{\bibinfo{volume}{102}},
  \bibinfo{pages}{017005} (\bibinfo{year}{2009}).

\bibitem[{\citenamefont{Vojta}(2009)}]{Vojta_stripe_review}
\bibinfo{author}{\bibfnamefont{M.}~\bibnamefont{Vojta}}, \bibinfo{journal}{Adv.
  Phys.} \textbf{\bibinfo{volume}{58}}, \bibinfo{pages}{699}
  (\bibinfo{year}{2009}).

\bibitem[{\citenamefont{Aji and Varma}(2007)}]{Aji-Varma_orbital_currents_PRL}
\bibinfo{author}{\bibfnamefont{V.}~\bibnamefont{Aji}} \bibnamefont{and}
  \bibinfo{author}{\bibfnamefont{C.~M.} \bibnamefont{Varma}},
  \bibinfo{journal}{Phys. Rev. Lett.} \textbf{\bibinfo{volume}{99}},
  \bibinfo{pages}{067003} (\bibinfo{year}{2007}).

\bibitem[{\citenamefont{Aji and Varma}(2009)}]{Aji-Varma_dissipative_XY}
\bibinfo{author}{\bibfnamefont{V.}~\bibnamefont{Aji}} \bibnamefont{and}
  \bibinfo{author}{\bibfnamefont{C.~M.} \bibnamefont{Varma}},
  \bibinfo{journal}{Phys. Rev. B} \textbf{\bibinfo{volume}{79}},
  \bibinfo{pages}{184501} (\bibinfo{year}{2009}).

\bibitem[{\citenamefont{Caldeira and Leggett}(1983)}]{Caldeira-Leggett}
\bibinfo{author}{\bibfnamefont{A.~O.} \bibnamefont{Caldeira}} \bibnamefont{and}
  \bibinfo{author}{\bibfnamefont{A.~J.} \bibnamefont{Leggett}},
  \bibinfo{journal}{Ann. Phys. (NY)} \textbf{\bibinfo{volume}{149}},
  \bibinfo{pages}{374 } (\bibinfo{year}{1983}).

\bibitem[{\citenamefont{Faulkner et~al.}(2010)\citenamefont{Faulkner, Iqbal,
  Liu, McGreevy, and Vegh}}]{Faulkner_strange_metal}
\bibinfo{author}{\bibfnamefont{T.}~\bibnamefont{Faulkner}},
  \bibinfo{author}{\bibfnamefont{N.}~\bibnamefont{Iqbal}},
  \bibinfo{author}{\bibfnamefont{H.}~\bibnamefont{Liu}},
  \bibinfo{author}{\bibfnamefont{J.}~\bibnamefont{McGreevy}}, \bibnamefont{and}
  \bibinfo{author}{\bibfnamefont{D.}~\bibnamefont{Vegh}},
  \bibinfo{journal}{Science} \textbf{\bibinfo{volume}{329}},
  \bibinfo{pages}{1043} (\bibinfo{year}{2010}).

\bibitem[{\citenamefont{Hertz}(1976)}]{Hertz_quantum_critical}
\bibinfo{author}{\bibfnamefont{J.~A.} \bibnamefont{Hertz}},
  \bibinfo{journal}{Phys. Rev. B} \textbf{\bibinfo{volume}{14}},
  \bibinfo{pages}{1165} (\bibinfo{year}{1976}).

\bibitem[{\citenamefont{Sperstad et~al.}(2010)\citenamefont{Sperstad, Stiansen,
  and Sudb\o{}}}]{Sperstad_dissipative}
\bibinfo{author}{\bibfnamefont{I.~B.} \bibnamefont{Sperstad}},
  \bibinfo{author}{\bibfnamefont{E.~B.} \bibnamefont{Stiansen}},
  \bibnamefont{and} \bibinfo{author}{\bibfnamefont{A.}~\bibnamefont{Sudb\o{}}},
  \bibinfo{journal}{Phys. Rev. B} \textbf{\bibinfo{volume}{81}},
  \bibinfo{pages}{104302} (\bibinfo{year}{2010}).

\bibitem[{\citenamefont{Chakravarty et~al.}(1986)\citenamefont{Chakravarty,
  Ingold, Kivelson, and Luther}}]{Chakravarty_dissipative_PT}
\bibinfo{author}{\bibfnamefont{S.}~\bibnamefont{Chakravarty}},
  \bibinfo{author}{\bibfnamefont{G.-L.} \bibnamefont{Ingold}},
  \bibinfo{author}{\bibfnamefont{S.}~\bibnamefont{Kivelson}}, \bibnamefont{and}
  \bibinfo{author}{\bibfnamefont{A.}~\bibnamefont{Luther}},
  \bibinfo{journal}{Phys. Rev. Lett.} \textbf{\bibinfo{volume}{56}},
  \bibinfo{pages}{2303} (\bibinfo{year}{1986}).

\bibitem[{\citenamefont{B{\o}rkje and
  Sudb{\o}}(2008)}]{Borkje_orbital_currents}
\bibinfo{author}{\bibfnamefont{K.}~\bibnamefont{B{\o}rkje}} \bibnamefont{and}
  \bibinfo{author}{\bibfnamefont{A.}~\bibnamefont{Sudb{\o}}},
  \bibinfo{journal}{Phys. Rev. B} \textbf{\bibinfo{volume}{77}},
  \bibinfo{eid}{092404} (\bibinfo{year}{2008}).

\bibitem[{\citenamefont{Sch\"{o}n and Zaikin}(1990)}]{Schon-Zaikin}
\bibinfo{author}{\bibfnamefont{G.}~\bibnamefont{Sch\"{o}n}} \bibnamefont{and}
  \bibinfo{author}{\bibfnamefont{A.~D.} \bibnamefont{Zaikin}},
  \bibinfo{journal}{Phys. Rep.} \textbf{\bibinfo{volume}{198}},
  \bibinfo{pages}{237 } (\bibinfo{year}{1990}).

\bibitem[{\citenamefont{Stiansen et~al.}(2011)\citenamefont{Stiansen, Sperstad,
  and Sudb\o{}}}]{Stiansen_compactness}
\bibinfo{author}{\bibfnamefont{E.~B.} \bibnamefont{Stiansen}},
  \bibinfo{author}{\bibfnamefont{I.~B.} \bibnamefont{Sperstad}},
  \bibnamefont{and} \bibinfo{author}{\bibfnamefont{A.}~\bibnamefont{Sudb\o{}}},
  \bibinfo{journal}{Phys. Rev. B} \textbf{\bibinfo{volume}{83}},
  \bibinfo{pages}{115134} (\bibinfo{year}{2011}).

\bibitem[{\citenamefont{Werner et~al.}(2005)\citenamefont{Werner, Troyer, and
  Sachdev}}]{Werner-Troyer-Sachdev_dissipative_XY_chain}
\bibinfo{author}{\bibfnamefont{P.}~\bibnamefont{Werner}},
  \bibinfo{author}{\bibfnamefont{M.}~\bibnamefont{Troyer}}, \bibnamefont{and}
  \bibinfo{author}{\bibfnamefont{S.}~\bibnamefont{Sachdev}},
  \bibinfo{journal}{J. Phys. Soc. Jpn.} \textbf{\bibinfo{volume}{74S}},
  \bibinfo{pages}{67} (\bibinfo{year}{2005}).

\bibitem[{\citenamefont{Hukushima and
  Nemoto}(1996)}]{Hukushima_parallel_tempering}
\bibinfo{author}{\bibfnamefont{K.}~\bibnamefont{Hukushima}} \bibnamefont{and}
  \bibinfo{author}{\bibfnamefont{K.}~\bibnamefont{Nemoto}},
  \bibinfo{journal}{J. Phys. Soc. Jpn.} \textbf{\bibinfo{volume}{65}},
  \bibinfo{pages}{1604} (\bibinfo{year}{1996}).

\bibitem[{\citenamefont{Matsumoto and Nishimura}(1998)}]{Mersenne_Twister}
\bibinfo{author}{\bibfnamefont{M.}~\bibnamefont{Matsumoto}} \bibnamefont{and}
  \bibinfo{author}{\bibfnamefont{T.}~\bibnamefont{Nishimura}},
  \bibinfo{journal}{ACM Trans. Model. Comput. Simul.}
  \textbf{\bibinfo{volume}{8}}, \bibinfo{pages}{3} (\bibinfo{year}{1998}).

\bibitem[{\citenamefont{Caselle and
  Hasenbusch}(1998)}]{Caselle-Hasenbusch_cubic_fixed_point}
\bibinfo{author}{\bibfnamefont{M.}~\bibnamefont{Caselle}} \bibnamefont{and}
  \bibinfo{author}{\bibfnamefont{M.}~\bibnamefont{Hasenbusch}},
  \bibinfo{journal}{J. Phys. A: Math. Gen.} \textbf{\bibinfo{volume}{31}},
  \bibinfo{pages}{4603} (\bibinfo{year}{1998}).

\bibitem[{(Footnote)}]{fotnote}
Here, we focused on the 
{\it perturbative} relevance of $h_4$. A substantial $h_4 > 1$ 
does seem to change the nature of the phase transition. The observation of diverging Binder cumulants 
$g \rightarrow -\infty$ points to a first-order phase transition both for the strongly anisotropic $XY$ case 
and for the $Z_4$ case.

\bibitem[{\citenamefont{Vojta}(2008)}]{Vojta_computing_QPT}
\bibinfo{author}{\bibfnamefont{T.}~\bibnamefont{Vojta}}, \bibinfo{journal}{Rev.
  Comput. Chem.} \textbf{\bibinfo{volume}{26}}, \bibinfo{pages}{167}
  (\bibinfo{year}{2008}).

\end{thebibliography}
\end{document}